\documentclass{svjour3}                     
\smartqed  
\usepackage{graphicx}
\usepackage{lineno}
\usepackage{setspace}
\usepackage[english]{babel}
\usepackage{latexsym}
\usepackage{amsmath}
\usepackage{amsfonts}
\usepackage{amssymb}
\usepackage[square,numbers]{natbib}
\usepackage[utf8]{inputenc}
\usepackage[T1]{fontenc}
\usepackage[active]{srcltx}
\usepackage{palatino}
\usepackage{psfrag}
\usepackage{epsfig}
\usepackage{color}
\usepackage{changebar}
\usepackage{fancyhdr}
\usepackage[active]{srcltx}
\usepackage[para]{footmisc}
\usepackage{pdflscape}
\usepackage{longtable}
\usepackage{upgreek}   
\usepackage{caption}
\usepackage[dvipsnames]{xcolor}
\usepackage{ulem} 
\usepackage{booktabs}
\usepackage{multirow}
\usepackage{tabularx}
\usepackage[misc]{ifsym} 
\usepackage[colorinlistoftodos]{todonotes}
\usepackage{subcaption}
\usepackage{gensymb}
\usepackage{placeins} 
\newcommand{\ApOne}[1]{\textcolor{black}{#1}}

\newcommand{\EqRefAp}[1]{Eq.~\eqref{#1}}

\newcommand{\FigRefAp}[1]{Fig.~\ref{#1}}
\newcommand{\TabRefAp}[1]{Tab.~\ref{#1}}

\renewcommand{\d}{{\,\rm  d}}

\newcommand{\T}[1]{{#1}^{\sf  T}}

\newcommand{\pd}[2]{\displaystyle\frac{\partial #1}{\partial #2}}
\newcommand{\pdf}[2]{\displaystyle{\partial #1}{/\partial #2}}

\newcommand{\grad}[1]{{\rm grad}\left( #1 \right)}
\renewcommand{\div}[1]{{\rm div }\left( #1 \right)}
\newcommand{\laplace}[1]{\bigtriangleup \left( #1 \right)}

\newcommand{\dev}[1]{{\rm dev }\left( #1 \right)}

\newcommand{\fabs}[1]{|| #1 ||}

\newcommand{\unit}[1]{{\rm #1}}
\newcommand{\lb}{\left(}
\newcommand{\rb}{\right)}

\newcommand{\f}[1]{\mbox{$ #1 $}}

\newcommand{\sty}[1]{\mbox{\boldmath $#1$}}
\newcommand{\styy}[1]{{\mathbb{#1}}}
\newcommand{\fa}{\sty{ a}}
\newcommand{\fb}{\sty{ b}}

\newcommand{\fe}{\sty{ e}}

\newcommand{\fg}{\sty{ g}}

\newcommand{\fn}{\sty{ n}}

\newcommand{\fq}{\sty{ q}}

\newcommand{\ft}{\sty{ t}}
\newcommand{\fu}{\sty{ u}}
\newcommand{\fv}{\sty{ v}}

\newcommand{\fx}{\sty{ x}}

\newcommand{\fzero}{\sty{ 0}}

\newcommand{\fB}{\sty{ B}}

\newcommand{\fD}{\sty{ D}}

\newcommand{\fI}{\sty{ I}}

\newcommand{\ffC}{\styy{ C}}

\newcommand{\falpha}{\mbox{\boldmath $\alpha$}}
\newcommand{\fsigma}{\mbox{\boldmath $\sigma$}}

\newcommand{\feps}{\mbox{\boldmath $\varepsilon $}}
\newcommand{\fphi}{\mbox{\boldmath $\phi $}}
\newcommand{\fkappa}{\mbox{\boldmath $\kappa $}}

\newcommand{\cVt}{{\cal V}_t}
\newcommand{\cVtBoundary}{\partial{\cal V}_t}
\newcommand{\cVtBoundaryNeum}{\partial{\cal V}_t^{\rm N}}
\newcommand{\cf}{cf.~}
\newcommand{\eg}{, e.g.,~}
\newcommand{\cfeg}{cf., e.g.,~}

\newcommand{\at}[2]{\left. #1 \right\vert_{#2}}

\newcommand{\MatTimeDerivative}[1]{\dot{#1}}
\newcommand{\elastic}{\rm{e}}
\newcommand{\plastic}{\rm{p}}

\newcommand{\entropyBulk}{\eta}

\newcommand{\temp}{T} 
\newcommand{\tempDot}{\dot{\temp}}
\newcommand{\tempFlux}{\fq}

\newcommand{\dissBulk}{\delta}
\newcommand{\energyFreeBulk}{\psi}

\newcommand{\energyInternalBulk}{e}

\newcommand{\StressCauchy}{\fsigma}
\newcommand{\StrainInf}{\feps}
\newcommand{\StrainInfElastic}{\feps^{\elastic}}
\newcommand{\StrainInfPlastic}{\feps^{\plastic}}

\newcommand{\StrainInfDot}{\MatTimeDerivative{\feps}}
\newcommand{\StrainInfPlasticDot}{\MatTimeDerivative{\feps}^{\plastic}}

\newcommand{\densityMass}{\rho}

\newcommand{\CoeffThermalExp}{\falpha}
\newcommand{\CoeffThermalExpScalar}{\alpha}

\newcommand{\HeatConductionTensor}{\fkappa}
\newcommand{\HeatConductionCoeff}{\kappa}
\newcommand{\ConsistencyParm}{\gamma}
\newcommand{\Yieldfun}{f}
\newcommand{\Stiffness}{\ffC}
\newcommand{\HeatCapacityStrainsRNTB}{\kappa_c}

\newcommand{\orderParamsVec}{\fphi}
\newcommand{\calF}{\mathcal{F}}
\newcommand{\freeEnergyPot}{f_{{\rm pot}}}
\newcommand{\freeEnergyGrad}{f_{{\rm grad}}}
\newcommand{\freeEnergyBulk}{\overline{f}_{{\rm bulk}}}
\newcommand{\numberPhases}{N}
\newcommand{\orderParams}[1]{\phi_{#1}}

\newcommand{\MobPhases}[1]{M_{#1}}
\newcommand{\vd}[2]{\displaystyle\frac{\delta #1}{\delta #2}}
\newcommand{\freeEnergyPFM}{f}

\newcommand{\InterfaceWidth}{\epsilon}
\newcommand{\StrainInfEigenCompon}[1]{\varepsilon_{#1}^{\ast}}
\newcommand{\StrainInfEigen}{\feps^{\ast}}

\newcommand{\StressYieldZero}{\sigma_{\rm{Y}}}
\newcommand{\StressBoundary}{\sigma_0}

\begin{document}
\raggedbottom

\title{The effect of compression shock heating in collision welding}


\author{G.C. Ganzenm\"uller$^{1,2}$ \and
        S. H\"utter$^3$ \and
        M. Reder$^{4,5}$ \and
        A. Prahs$^4$ \and
        D. Schneider$^{4,6}$ \and
        B. Nestler$^{4,5,6}$ \and
        T. Halle$^{3}$ \and
        S. Hiermaier$^{1,2}$        
}

\authorrunning{G.C. Ganzenm\"uller \textit{et al.}} 


\institute{$^1$ Albert-Ludwigs Universität Freiburg, INATECH, 79110 Freiburg i. Br., Germany\\
           $^2$ Fraunhofer Ernst-Mach-Institute for High-Speed Dynamics, EMI, 79104 Freiburg i. Br., Germany
           \at \email{georg.ganzenmueller@inatech.uni-freiburg.de}
           \\
           $^3$ Institute of Materials and Joining Technology, Otto-von-Guericke University, 39106 Magdeburg, Germany
           \\
           $^4$ Institute for Applied Materials – Microstructure Modelling and Simulation, Karlsruhe Institute of Technology (KIT), Straße am Forum 7, Karlsruhe, 76131, Germany        
           \\
           $^5$ Institute of Digital Materials Science (IDM), Karlsruhe University of Applied Sciences, Moltkestraße 30, Karlsruhe, 76133, Baden-Württemberg, Germany
           \\
           $^6$ Institute of Nanotechnology (INT), Karlsruhe Institute of Technology (KIT), Hermann-von-Helmholtz-Platz 1, Eggenstein-Leopoldshafen, 76344, Baden-Württemberg, Germany          
}

\date{Received: date / Accepted: date}

\maketitle


\begin{abstract}
This work discusses the origin of temperature rise during the collision welding
process. The different physical irreversible and reversible mechanisms which act
as heat sources are described: isentropic compression work, shock dissipation,
plasticity, and phase transitions. The temperature increase due to these effects
is quantified in a continuum mechanics approach, and compared to predictions of
atomistic molecular dynamics simulations. Focusing on a single impact scenario
of 1100 aluminium at 700 m/s, our results indicate that shock heating and
plastic work only effect a temperature rise of 100 K, and that the effects of
phase change are not significant. This temperature rise cannot explain welding.
In consequence, the relevant mechanism which effects bonding in collision
welding must be due to the jet, which is only formed at oblique impact angles.
\end{abstract}

%
\section{Introduction}
\label{sec:intro}
%



Collision welding, also known as impact welding or high-velocity impact welding, is a solid-state welding process that involves the joining of two metal components through a high-speed collision. In this process, two metal surfaces are brought into contact under a high velocity and pressure, resulting in a localized plastic deformation and the formation of a strong metallurgical bond. \cite{wang_high-velocity_2019}

Collision welding offers several advantages over traditional welding methods. It does not require the use of filler materials or external heat sources, making it a cost-effective and efficient process. Additionally, collision welding can join dissimilar metals that are difficult to weld using conventional methods. \cite{Kacar_2004}

During collision welding, the two metal parts need to collide at a certain angle, typically between 5\textdegree to 10\textdegree. During this oblique impact, most of the kinetic energy is transformed rapidly into pressure. This process is termed shock compression. If the collision angle is large enough, the highly pressurized material can escape towards the side where no contact has been made yet, forming a supersonic jet of material, which exhibits features of a periodic hydrodynamic instability, and is thought to be very hot, in excess of 1000 K \cite{Niessen_2020}. One possible explanation as to why collision welding works so well is that the jet interacts with the metals surfaces that have not been in contact yet, and cleans these in an abrasive fashion, facilitating cold welding \cite{bataev_towards_2019}.

However, it is currently unclear if this proposed mechanism is indeed applicable \cite{Niessen_2018a}, and what the role of shock compression is. Shock compression refers to the rapid increase in pressure and temperature experienced by a material due to the application of a sudden and intense mechanical shock wave. When a material is subjected to shock compression, it undergoes rapid compression and heating, leading to significant changes in its physical and mechanical properties. The high-pressure and high-temperature conditions induced by shock compression can cause structural changes and phase transformations.

To further understand the role of the jet, it is important to first quantify the effects of shock compression, in particular the associated temperature rise. It needs to be clarified if this temperature rise alone can lead to melting and bonding of two metal parts, or whether an additional mechanism -- such as the jet -- is necessarily required to achieve welding. The shock wave was investigated in \cite{williams_analysis_2018}, with a particular emphasis on pressure and density changes, but the effects of temperature due to shock heating were not addressed. This work seeks to fill in this gap.

To this end, we limit ourselves to studying only the normal impact of 1100 Aluminium at
700 m/s, a typical velocity for collision welding \cite{HAHN20168}. We consider three possible sources of heat: adiabatic shock
compression, dissipation by plastic deformation, and generation of defects on a
microstructural scale. We first establish an analytical framework to predict
temperature increase from a pure continuum mechanics perspective, taking into
account only the thermodynamics of shock compression based on the equation of
state and a constant yield stress of the material. This analysis is augmented
with a model that estimates the temperature contribution of defects generation.
The latter model takes into account that a defect has different stiffness and
strength compared to the surrounding materials, leading to dissipation when the
associated stress field is integrated along the deformation path.

The prediction of the continuum model is compared against Molecular Dynamics
(MD) simulations. In contrast to the continuum model, where isotropic material
properties are assumed, MD requires a description of the actual atomistic
structure. It is impractical, due to limited computer resources, to
simulate a statistically large enough collection of single crystal grains to
achieve an isotropic distribution of material properties. Therefore, the collision process of individual single crystals along different lattice orientations are simulated. The resulting MD ensemble temperatures agree with the prediction of the continuum model, both in the shocked state, and after shock release.

The remainder of this article is organized as follows: We first introduce the shock compression analysis and its associated temperature increase. We then consider the role of defects and estimate their temperature contribution. Finally, the predictions from the theoretical analysis are compared to MD simulations, and the results are discussed.

%
\section{Macroscopic perspective on heating during shock compression}
\label{sec:EOS}
%

This section describes how shock compression leads to temperature increase from a purely macroscopic perspective, considering only the thermodynamics of the equation of state and the dissipative contribution due to plastic work. No effects due to phase changes are taken into account. The discussion of the contribution of phase change to temperature increase is deferred to Sec.~\ref{sec:phase_changes}. 

\subsection{Hydrodynamic Adiabatic Heating during shock compression}
We begin the description of the temperature evolution by describing the shock compression of a purely hydrodynamic
material. This is an inviscid fluid whose pressure is defined by an equation of
state (EOS), which depends only on volumetric compression ratio and internal
energy.

\begin{figure}
	\centering
	\includegraphics[width=8cm]{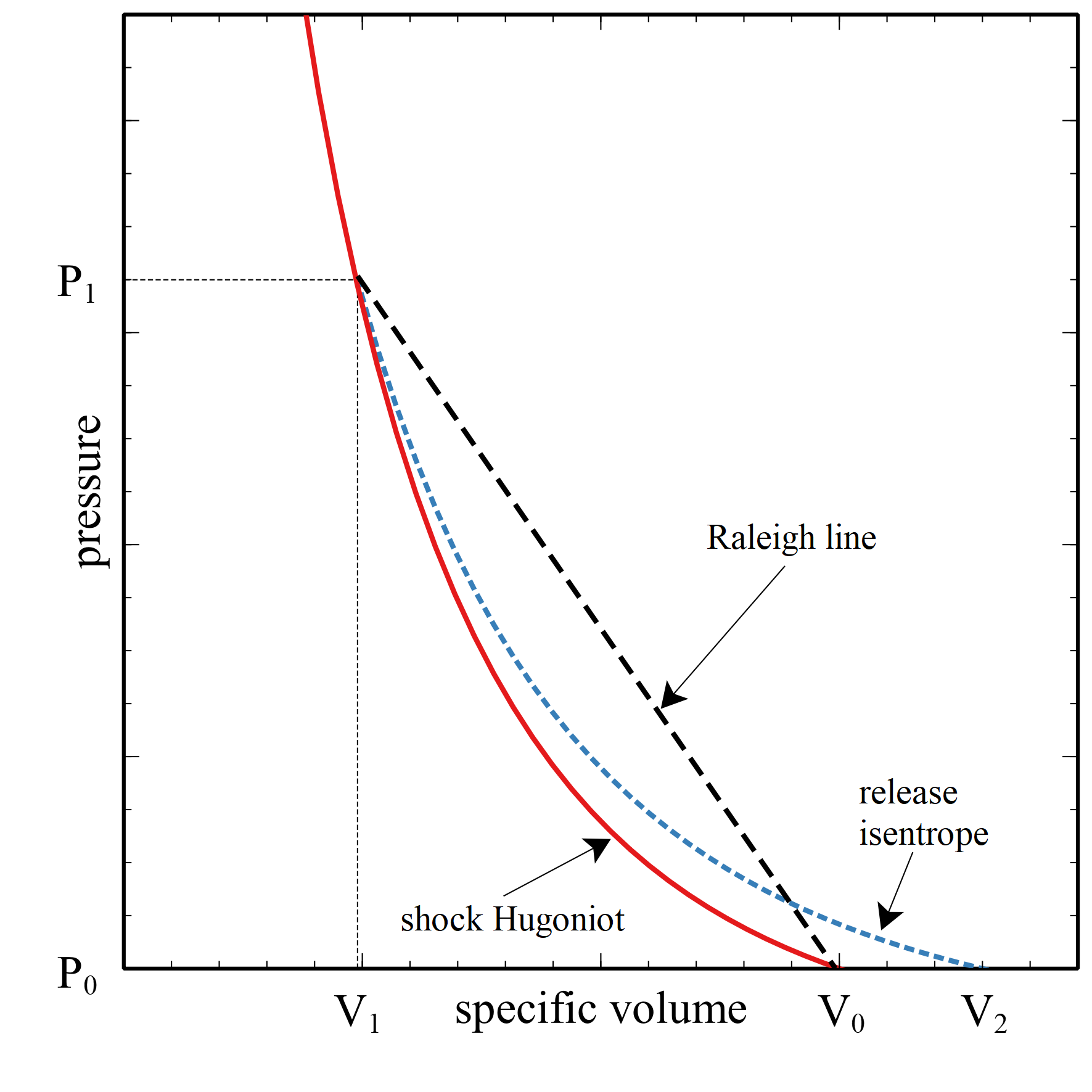}
	\caption{Shock compression is considered an instantaneous process from
	 $\left(V_0, P_0\right)$ along the straight Raleigh line to $\left(V_1,
	 P_1\right)$. Unloading occurs along the curved Hugoniot back to the
	 released state $\left(V_2, P_2\right)$}
	\label{fig:shock_compression_release}
\end{figure}

It is well known in the literature that shock compression leads to temperature
increase. Phenomenologically, this may be understood by considering the
irreversibility of the compression, see
Fig.~\ref{fig:shock_compression_release}. Shock loading is assumed to be
instantaneous from the initial rest state at ambient conditions, corresponding
to $\left(V_0, P_0\right)$, to the compressed state $\left(V_1, P_1\right)$.
Here, $V=1/\rho$ denotes specific volume, which is the inverse of mass
density $\rho$. $P$ refers to pressure, and the subscripts denote different
compression states. Because shock loading is assumed to happen instantaneously,
the loading path follows the straight \textit{Raleigh} line. Unloading from the
shocked state to the released state at ambient pressure, in contrast, is not
forced and happens along the natural path describe by the equation of state,
which is commonly termed the \textit{Hugoniot} path. Typically, the Hugoniot is a convex
function rather than being a straight line. As loading and unloading paths are
different, they form a hysteresis loop with the enclosed area corresponding to
dissipated energy, which contributes to an elevated temperature of the material
even after the shock pressure has been released.

A procedure for calculating the temperature in the shocked state, $\left(V_1,
P_1\right)$, has been devised by Walsh and Christian in 1955
\cite{walsh_equation_1955}. As the full derivation is lengthy and is
available in standard textbooks, we only quote the most important steps. The
derivation starts from the first law of thermodynamics, $\mathrm{d}E = \delta Q
- \delta W$, and assumes that all work is against volume. With the definition of
temperature $T$ in terms of entropy and heat change, $\delta Q / T = \mathrm{d}
S$, the starting point reads:

\begin{equation}
    \mathrm{d} E = T \mathrm{d} S - P \mathrm{d} V
\end{equation}
Manipulation using Maxwell's relations, the definition of the heat capacity,\\
$\left(\partial E / \partial T\right)_V = C_V$, and the definition of the
Grüneisen parameter $\gamma = \left(\partial P / \partial T\right)_V V$ results
in 

\begin{equation}
    \mathrm{d}E = C_v \mathrm{d} T + T \frac{\gamma}{V} C_v \mathrm{d} V - P \mathrm{d} V \label{eq:dE}
\end{equation}
The latter describes the continuous change of internal energy as a function of
change in volume. Shock loading changes the volume from the initial state $V_0$
to the shocked state $V_1$. For this process, the change in energy is known from
the Rankine-Hugoniot jump conditions, which are the consequence of conservation laws across the
shock. In Fig.~\ref{fig:shock_compression_release}, this jump is visualized by
the Raleigh line.

\begin{equation}
\Delta E = E_1 - E_0 = \frac{1}{2} (P_1 + P_0) (V_0 - V_1) \label{eq:DeltaE}
\end{equation}
Combining Eqns.~\ref{eq:dE} and \ref{eq:DeltaE} yields a differential equation
for the temperature evolution when compressing from $V_0$ to $V_1$. The solution
to this equation is

\begin{equation}
	T_1 = T_0 e^{b (V_0-V_1)} + \frac{P_1 (V_0 - V_1)}{2 C_v} + \frac{1}{2 C_v} e^{-b V_1} \int\limits_{V_0}^{V_1} P(V) e^{b V} \left[ 2 - b (V_0-V)\right] \mathrm{d}V \label{eq:T1}
\end{equation}
with $b=\gamma_0 / V_0$. The integral has to be evaluated numerically, and an
specific EOS is needed. In the shock context, it is natural to choose the shock
Hugoniot EOS, which uses only two parameters, that are readily available from
experiment:

\begin{eqnarray}
	P(V) &=& \frac{C^2 (V_0 - V)}{[V_0 - S(V_0-V)]^2}\\
	P(U_p) &=& \frac{1}{V_0} \left( C U_p+ S U_p^2 \right)
\end{eqnarray}
$C$ and $S$ are the intercept and slope when the pressure front velocity is
plotted against the impact velocity in a shock experiment. Eq.~\ref{eq:T1} thus
yields the temperature in the shocked state. It is also interesting to compute
the temperature after releasing the shock pressure. It is unknown which exact
thermodynamic path to follow here, but one can argue that it is both adiabatic
and reversible, thus isentropic: The unloading process happens too quick
to exchange significant amounts of heat with the environment, and the material
expands along its EOS. Thus, one may invoke Eq.~\ref{eq:dE} and set
$\mathrm{d}S=0$, yielding

\begin{equation}
	\frac{\mathrm{d}T}{T} = -\frac{\gamma}{V} \mathrm{d}V
\end{equation}
Assuming that the Grüneisen ratio, $\gamma/V = \gamma_0 / V_0$ is constant, one
obtains by integrating from $\left(V_1,P_1\right)$ to the released state
$\left(V_2,P_2\right)$:

\begin{equation}
	T_2 = T_1 e^{b \left(V_1-V_2\right)} \label{eq:T2}
\end{equation}
Eq.~\ref{eq:T1} and \ref{eq:T2} yield the purely hydrodynamic temperature changes
for shock compression and release, considering only changes in volume. This
would be applicable to a fluid without viscosity. In the case of metals,
however, two more relevant heat sources need to be addressed, plasticity and
phase changes. Here, we incorporate only the contribution due to plastic work, and defer the effect of phase changes to Sec.~\ref{sec:phase_changes}.
\subsection{Plastic work}
Shock compression considered to be a fast uniaxial deformation, i.e., the
material volume is changed along a non-isotropic path. This deformation leads to
shear strains, which will be expressed as equivalent uniaxial strain, $\varepsilon_{eq}$, in the
following. To this end, we note that the determinant of the deformation gradient
\ApOne{$\mathbf{F}$}, $J$,  describes the volumetric deformation as $\mathrm{d}V = J \mathrm{d}V_0$, so

\begin{equation}
	J = \frac{V_1}{V_0}
\end{equation}
For uniaxial deformation by compression along one Cartesian direction $x$, $\mathbf{F}$ is diagonal.

\begin{equation}
	\mathbf{F} = \begin{bmatrix}
		F_{xx} & 0 & 0 \\
		0 & 1 & 0 \\
		0 & 0 & 1 \\
		\end{bmatrix}
\end{equation}
We thus have $\mathrm{det} \mathbf{F} = F_{xx} = \frac{V_1}{V_0}$, which fully determines $\mathbf{F}$.
The Green-Lagrange strain is obtained as $\mathbf{E} = \frac{1}{2}
\left(\mathbf{F}^T\mathbf{F} -I\right)$. The equivalent uniaxial strain is
obtained as 

\begin{equation}
	\varepsilon_{eq} = \sqrt{\frac{2}{3}} \left|\mathbf{E}_{\mathrm{dev}}\right|
\end{equation}
where $\mathbf{E}_{\mathrm{dev}}$ is the deviatoric part of the Green-Lagrange strain tensor.
%
%
%

Plastic work per unit initial volume can thus be obtained by integrating the yield stress~\ApOne{\f{\StressYieldZero}} along the deformation path,

\begin{equation}
	\Delta Q_{plastic} / V_0 = \int\limits_{0}^{\varepsilon_{eq}}  \ApOne{\f{\StressYieldZero}} \mathrm{d}\varepsilon
\end{equation}
Note that plastic work is performed twice, once on loading, and again on
unloading back to $V_2 \approx V_0$. Assuming that both yield stress and heat capacity are constant, the temperature
increase due to plastic work is:

\begin{equation}
	\Delta T_{plastic} = \frac{\beta}{c_V} \varepsilon_{eq} \ApOne{\f{\StressYieldZero}} V_0
\end{equation}
Here, $\beta$ is the Taylor-Quinney coefficient, which specifies the fraction of
plastic work converted into heat. A typical assumption is that $\beta=0.9$, and
the remaining work is stored in microstructure changes.

\begin{table}[] \label{tab:Tshock}
	\caption{Equation of state and material strength parameters for aluminium. Data taken from references \cite{Steinberg_1996, Mitchell_1981}
	}
	\begin{tabular}{lll}
	parameter              			& value 				& units \\
	\hline
	Mie-Grüneisen $\gamma$ 			& 1.97  				& -     \\
	shock Hugoniot intercept $C$ 	& 5240  				& m/s   \\
	shock Hugoniot slope $S$ 		& 1.4					& -		\\
	specific volume $V_0$			& 3.70$\times 10^-4$	& m$^3$/kg \\
	heat capacity $c_V$				& 885					& J/(kg K) \\
	yield stress $\ApOne{\f{\StressYieldZero}}$			& 80					& MPa
	\end{tabular}
\end{table}

\begin{table}[]
	\caption{Temperatures due to shock compression, plastic dissipation, and
		subsequent release to zero pressure. Note that the initial system
		temperature is assumed as $T_0=300$ K. $T_{\mathrm{shock}}=T_1+\Delta T_{\mathrm{plastic}}$ is the
		temperature in the shocked state, $T_{\mathrm{release}}=T_2+2\Delta T_{\mathrm{plastic}}$ is the
		temperature after release.}
	\label{tab:cont_predict_temp}
	\begin{tabular}{lllllll}
	impact velocity & $V_1/V_0$ & $P_1$ & $\varepsilon_{eq}$ & $\Delta T_{\mathrm{plastic}}$ & $T_{\mathrm{shock}}$ & $T_{\mathrm{release}}$\\
	m/s                       & -  & GPa  & -  & K  & K & K\\
	\hline
	700                       & 0.89  & 11.8  & 0.11  & 3.5  & 403.1 & 330.4
	\end{tabular}
\end{table}

We summarize this section by noting that macroscopic analysis, based on the thermodynamics of the EOS and plastic work, leads to a temperature increase of 103.1 K in the shocked state. After cooling due to isentropic release, the remaining temperature increase compared to the pre-shocked state is 30.4 K.

%
\FloatBarrier
\section{Continuum mechanics perspective on heating due to phase change.}
\label{sec:phase_changes}
%
\ApOne{
The main objective of this section is to discuss the role of a diffusionless, solid-solid phase transformation as a heat source.
To this end a setup is considered which is closely related to early analytical and numerical attempts \cite{Leblond1986} for simulating the Greenwood-Johnson effect~\cite{Greenwood1965}: A representative volume element consisting of a distorted elastoplastic inlusion embedded in an elastoplastic matrix under external load. 
The distortion mimics the difference in mass density between the matrix and inclusion and is modeled by non-vanishing eigenstrains inside the inclusion.
The yield stresses of matrix and inclusion differ with the matrix being the weaker phase. 
Under the applied load, the inclusion grows at the expense of the matrix.
This displacive, solid-solid phase transition is representative for a change of the underlying crystal lattice from face-centered cubic to body-centered tetragonal, as in the case of martensitic phase transformation. 
The difference in the yield stress contributes to the driving-force of the phase-transition. 
Therefore, while the yield stress of the inclusion is considered as in the previous section, the yield stress of the matrix is assumed to be a quarter of this, resulting in a weaker matrix phase. 
For the simulations a thermomechanically coupled theory (TCT) is considered, which is discussed in detail by~\citet{Prahs2023} in the context of the multiphase-field method. 
Subsequently, a brief summary of the TCT is provided, followed by the simulation results. 
}
%
%
%
%
\ApOne{A} direct tensor notation is used\ApOne{:} Tensors of first and second order are 
\ApOne{written as}
~\f{\fa} and~\f{\fB}.

\subsection{Thermomechanically coupled theory}
\paragraph{Balance equations}
Considering the current configuration of a material volume, the balance of total energy reads

\begin{align}
  \frac{\d}{\d t} \int_{\cVt} \densityMass \lb e + \frac{1}{2} \fv \cdot \fv \rb \d v 
  = \int_{\cVt} \densityMass \lb \fb \cdot \fv + r \rb \d v + \int_{\partial \cVt} \lb \ft \cdot \fv + h \rb \d a,
\label{eq:BalanceOfEnergy_Cauchy}
\end{align}
with the mass density~\f{\densityMass}, the specific free energy~\f{\energyInternalBulk}, the velocity field~\f{\fv}, the body force~\f{\fb}, the specific heat supply~\f{r}, the traction vector~\f{\ft} and the heat flux~\f{h}. 
%
The Cauchy stress tensor~\f{\StressCauchy} and the heat flux vector~\f{\tempFlux}\ApOne{, given by~\f{\ft = \StressCauchy \fn} and~\f{h = -\tempFlux \cdot \fn}, }
%
%
as well as the balance of mass\ApOne{~\f{\dot{\densityMass} + \densityMass \div{\fv} = 0}}, linear\ApOne{~momentum~\f{\densityMass \fa  = \div{\fsigma} + \densityMass \fb},} and angular momentum\ApOne{~\f{\fsigma = \T{\fsigma}}}
are obtained by means of invariance of the balance of total energy with respect to a change of observer, \cfeg\citet{Prahs2019,Prahs2019a} for a detailed discussion with respect to regular \ApOne{and} singular points. 
The subsequent considerations are limited to small deformations.
Thus, the density is considered to be a constant and small strains are considered, leading to~\f{\fD = \StrainInfDot}, \cfeg\citet[pp.30-32]{Maugin1992}. 
\ApOne{Moreover, the quasi-static case and vanishing body forces are considered, i.e.,~\f{\fa = \fzero} and~\f{\fb = \fzero} hold true.}
Taking into account \ApOne{the definition of the Cachy stress tensor and the heat flux, as well as the balance of mass, linear, and angular momentum}, the localization of~\EqRefAp{eq:BalanceOfEnergy_Cauchy} yields the balance of internal energy, reading
\begin{align}
    \densityMass \dot{e} = \densityMass r + \StressCauchy \cdot \StrainInfDot -\div{\tempFlux}.
    \label{eq:BalaceInternalEnergy_Cauchy}
\end{align}

\paragraph{\ApOne{Summary of constitutive equations and heat conduction}}
As a special form of the dissipation inequality, the Clausius--Duhem inequality is considered.
\ApOne{Taking into account}~\f{\fg = \grad{\temp}} \ApOne{and} the Legendre transformation~\f{\energyFreeBulk = \energyInternalBulk - \temp \eta}, cf., e,g.,~\cite{Beegle1974}, as well as taking into account~\EqRefAp{eq:BalaceInternalEnergy_Cauchy}, the Clausius--Duhem inequality reads
\begin{equation}
 \densityMass \dissBulk = \StressCauchy \cdot \StrainInfDot - \densityMass \dot{\energyFreeBulk} - \densityMass \dot{\temp}\entropyBulk - \frac{1}{\temp}\tempFlux \cdot \fg \geq 0,
 \label{equation:CDU_WithInternalEnergy}
\end{equation}
\cfeg\citet[p.152]{Bertram2005}. 
Here, the specific dissipation is referred to as~\f{\dissBulk}, the specific free energy as~\f{\energyFreeBulk}, and the specific entropy as~\f{\entropyBulk}. 
In the context of small deformations, an additive decomposition of the small strain tensor into an elastic~\f{\StrainInfElastic}, a plastic~\f{\StrainInfPlastic}, and a thermal contribution~\f{\CoeffThermalExp \Delta \temp}\ApOne{, with~\f{\Delta \temp = \temp - \temp_0}, as well as a contribution related to an eigenstrain~\f{\StrainInfEigen}} is assumed. 
Here, \f{\temp_0} refers to the referential temperature, and~\f{\CoeffThermalExp} to the coefficient of thermal expansion.
Subsequently, the free energy is considered to depend on the total strain~\f{\StrainInf}, the plastic strain~\f{\StrainInfPlastic}, the eigenstrain~\f{\StrainInfEigen,} and the temperature~\f{\temp}, reading\ApOne{~\f{\energyFreeBulk = \energyFreeBulk(\StrainInf, \StrainInfPlastic, \StrainInfEigen,  \temp)}.}
%
The exploitation of~\EqRefAp{equation:CDU_WithInternalEnergy} according to the Coleman-Noll procedure,\cfeg\cite{Coleman1967}, leads to the potential relations for the Cauchy stress\ApOne{~\f{\StressCauchy = \rho \pdf{\energyFreeBulk}{\StrainInf}}} and the entropy \ApOne{\f{\entropyBulk = -\pdf{\energyFreeBulk}{\temp}}}.
%
Within this section, the considered specific free energy reads
\begin{multline}
    \energyFreeBulk =
    \frac{1}{2 \densityMass} \lb \StrainInf - \StrainInfPlastic - \CoeffThermalExpScalar \Delta \temp \fI  - \StrainInfEigen \rb \cdot \lb \Stiffness \left[ \StrainInf - \StrainInfPlastic - \CoeffThermalExpScalar \Delta \temp \fI - \StrainInfEigen \right]   \rb 
     \\
    - \frac{1}{2 \densityMass} \lb \CoeffThermalExpScalar \Delta \temp \fI \rb \cdot \lb \CoeffThermalExpScalar \Delta \temp \fI \rb - \HeatCapacityStrainsRNTB\lb \temp \ln{\frac{\temp}{\temp_0}} - \Delta \temp \rb,
    \label{eq:SpecificFreeEnergy}
\end{multline}
\cfeg\cite{Neumann2016}.
Here,~\f{\CoeffThermalExpScalar} denotes the scalar-valued coefficient of thermal expansion,~\f{\fI} the identity tensor of second order, accounting for an isotropic thermal expansion, and~\f{\HeatCapacityStrainsRNTB} the heat capacity. 
%
\ApOne{
The common form of Fourier's law is assumed \cite{Bertram2005} and isotropic heat conduction is considered, for simplicity. 
Subsequently, Mises plasticity with associated flow rule is considered, \cfeg\citet[p.83]{Simo2008}. 
Hooke's and Fourier's law as well as the yield criterion and the flow rule are provided in~\TabRefAp{tab:SummaryOfConstEqs}.}
Within this section, an elastic perfectly plastic material behavior is considered, i.e., hardening is neglected, for brevity, \cfeg\citet[Fig.3.28]{Lemaitre2000}.
\renewcommand{\arraystretch}{1.75}
\begin{table}[htp!]
    \centering
\begin{tabular}{||l|l||}
Hooke's law    & ~\f{\StressCauchy = \Stiffness \left[ \StrainInf - \StrainInfPlastic - \CoeffThermalExpScalar \Delta \temp \fI - \StrainInfEigen \right]}, \, \f{\Delta \temp = \temp - \temp_0}  \\ \hline
Mises yield criterion & \f{\Yieldfun = \fabs{ \dev{\StressCauchy} } - \sqrt{\frac{2}{3}}\StressYieldZero \leq 0} \\ \hline
Flow rule & \f{\StrainInfPlasticDot = \ConsistencyParm \pd{ \Yieldfun }{\dev{\StressCauchy}} = \ConsistencyParm \frac{ \dev{\StressCauchy} }{\fabs{\dev{\StressCauchy}}}}, \, \f{\ConsistencyParm \geq 0} \\ \hline
Fourier's law & \f{\tempFlux = -\HeatConductionTensor \fg}, \, \f{\HeatConductionTensor = \HeatConductionCoeff \fI} \, \f{\fg = \grad{\temp}} 
%
\end{tabular}
    \caption{
    \ApOne{
    Here,~\f{\Stiffness} denotes the stiffness tensor of fourth order,~\f{\dev{\StressCauchy}} the deviator of the Cauchy stress tensor,~\f{\StressYieldZero} the initial yield stress, and~\f{\ConsistencyParm} the consistency parameter. 
    An associated plasticity is considered, establishing a relation between the yield criterion and the flow rule, \cfeg~\citet[p.195]{Lemaitre2000}. 
    Moreover,~\f{\HeatConductionTensor} denotes the positive definite heat conduction tensor, which is considered to be isotropic with the constant coefficient~\f{\HeatConductionCoeff}.
    }
    }
    \label{tab:SummaryOfConstEqs}
\end{table}
\ApOne{
Taking into account the Legendre transformation and the potential relations, as well as neglecting the heat supply, i.e.~\f{r = 0}, the balance of internal energy, \cf~\EqRefAp{eq:BalaceInternalEnergy_Cauchy}, leads to the following form of the heat conduction equation
}
\begin{align}
    \densityMass \HeatCapacityStrainsRNTB \tempDot = \HeatConductionCoeff \laplace{\temp} + \StressCauchy\cdot\StrainInfPlasticDot + \temp \pd{\StressCauchy}{\temp} \cdot \lb \StrainInfDot - \StrainInfPlasticDot \rb,
    \label{equation:HeatConduction_V7}
\end{align}
\ApOne{similar to~\citet[p.67]{Lemaitre2000}.}
\ApOne{The TCT, considered in the work at hand, accounts for the interactions between the balance of linear momentum and the heat conduction.}
Consequently, \ApOne{it} is in contrast to the assumptions and consequences of a purely mechanical theory as discussed by~\citet{Prahs2021}.

\paragraph{Phase-field method} The evolution of the phases is modelled by means of the multi-phase field method, \cfeg\citet{Steinbach1996, Steinbach1999, Nestler2005, Moelans2008}. 
To this end, the following free energy functional is considered
\begin{align}
    \calF \left[ \orderParamsVec, \nabla \orderParamsVec, \fu \right] = \int_{\cVt} \freeEnergyGrad
     + \freeEnergyPot
     + \freeEnergyBulk 
     \d v = \int_{\cVt} \freeEnergyPFM \d v
     \label{eq:FreeEnergyFunctional}
\end{align}
\cfeg\citet{Herrmann2018}, where~\f{\calF} denotes the free energy functional. 
In the work at hand,~\f{\orderParamsVec = \left\{ \orderParams{1}, \ldots, \orderParams{\numberPhases} \right\}} refers to the vector of order parameters, \f{\nabla \orderParamsVec} to its gradient, and~\f{\fu} to the displacement field.
Regarding the order parameters, the summation constraint~\f{\sum_{\alpha} \orderParams{\alpha} = 1} has to be fulfilled, \cfeg\citet[p.2]{Nestler2005}.
The free energy density~\f{\freeEnergyPFM} consists of the contributions~\f{\freeEnergyGrad} and~\f{\freeEnergyPot}, arising from the regularization of a sharp interface in a diffuse interface context, as well as the contribution~\f{\freeEnergyBulk} accounting for the free energy density of the bulk material.
A detailed discussion regarding the contributions~\f{\freeEnergyGrad} and~\f{\freeEnergyPot} is provided\eg by~\citet{Nestler2005}.
The bulk contribution~\f{\freeEnergyBulk} is given by the interpolation
\begin{align}
    \freeEnergyBulk &= \sum_{\alpha} \orderParams{\alpha} \rho_{\alpha} \energyFreeBulk_{\alpha},
\end{align}
where~\f{\energyFreeBulk_{\alpha}} refers to the specific free energy according to~\EqRefAp{eq:SpecificFreeEnergy} with respect to phase~\f{\alpha}. 
The evolution equation for phase~\f{\alpha} is obtained by
\begin{align}
    \pd{\orderParams{\alpha}}{t} = -\frac{1}{\InterfaceWidth\numberPhases} \sum\nolimits_{\beta \neq \alpha}^\numberPhases \MobPhases{\alpha \beta} \left( \vd{\freeEnergyPFM}{\orderParams{\alpha}} - \vd{\freeEnergyPFM}{\orderParams{\beta}}  \right), \quad \vd{\freeEnergyPFM}{\orderParams{\alpha}} =  \lb \pd{f}{\orderParams{\alpha}} - \div{\pd{f}{\nabla \orderParams{\alpha}}} \rb
\end{align}
where~\f{\MobPhases{\alpha \beta}} denotes the mobility of the diffuse interface between phase~\f{\alpha} and~\f{\beta}, and~\f{\numberPhases} the amount of locally present phases.
The parameter \f{\InterfaceWidth} is proportional to the interface width. 
Moreover, \ApOne{\f{ \pdf{\freeEnergyGrad}{\nabla \orderParams{\alpha}}\cdot \fn = 0, \forall \fx \in \cVtBoundaryNeum }}
holds true with~\f{\cVtBoundaryNeum} denoting the Neumann boundary of~\f{\cVtBoundary.}

\subsection{Simulations}
\begin{figure}[htp!]
    \centering
    \includegraphics[scale=0.45]{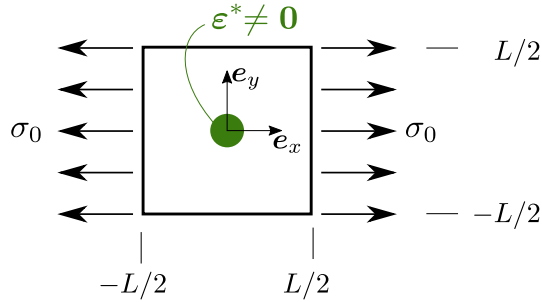}
    \caption{Illustration of considered boundary conditions\ApOne{, \cf also~\cite{Prahs2023}: An elastoplastic inlcusion is subjected to eigenstrains with \f{\StrainInfEigenCompon{xx} = \StrainInfEigenCompon{yy} = 0.002} and~\f{\StrainInfEigenCompon{zz} = \StrainInfEigenCompon{xy} = \StrainInfEigenCompon{xz} = \StrainInfEigenCompon{yz} = 0} is embedded in an elastoplastic matrix. 
    A plane strain setup is considered and non-vanishing Neumann boundary conditions on the left and right sides of the inclusion represent the load.}
    }
    \label{fig:IllustrationBoundaryConds}
\end{figure}

\paragraph{Boundary condition} 
Subsequently, an elastoplastic inclusion, subjected to eigenstrains, is located within an elastoplastic matrix. 
Both, the matrix and the inclusion, exhibit \ApOne{an isotropic} elastic material behavior, \ApOne{using Young's modulus and Poisson's ratio} according to~\TabRefAp{tab:md_material_params}, mimicking \ApOne{1100 aluminium}. 
In addition to~\TabRefAp{tab:md_material_params}, for both the matrix and the inclusion, the \ApOne{heat capacity is chosen as~\f{\HeatCapacityStrainsRNTB = 900~\unit{m^2/\left(s^2 K\right)}}, the heat conduction coefficient as~\f{\HeatConductionCoeff = 115~\unit{kg\, m /(s^3 K)}}, and the coefficient of thermal expansion as~\f{\CoeffThermalExpScalar = 2.38\times10^{-5}~\unit{1/K}}. 
The yield stress of the matrix is chosen as~\f{\StressYieldZero = 20~\unit{MPa}} while that of the inclusion is chosen considered to be~\f{\StressYieldZero = 80~\unit{MPa}}. 
Therefore, the yield stress of the inclusion is significantly larger than that of the weaker matrix.
} 
The boundary conditions for the simulations are illustrated by~\FigRefAp{fig:IllustrationBoundaryConds}. 
A plane strain set-up is considered with the Neumann boundary conditions,
\begin{align}
    \overline{\ft}\left( L/2, y, z \right) = \StressBoundary\fe_x, \quad  \overline{\ft}\left( -L/2, y, z \right) = -\StressBoundary\fe_x, \quad \overline{\ft}\left( x, \ApOne{\pm} L/2, z \right) = \fzero
    \label{eq:NeumannBCs1}
\end{align}
where~\f{L} denotes the width and height of the quadratic simulation domain and \f{\StressBoundary = 8 \unit{MPa}} is the stress applied at the boundary.
In addition, the Dirichlet boundary conditions~\ApOne{\f{u_z\left( x, y, \pm B/2 \right) = 0}.}
are applied regarding the~\f{z}-component of the displacement field~\f{\fu}.
The eigenstrains~\f{\StrainInfEigen} are considered to be~\ApOne{\f{\StrainInfEigenCompon{xx} = \StrainInfEigenCompon{yy} = 0.002} and~\f{\StrainInfEigenCompon{zz} = \StrainInfEigenCompon{xy} = \StrainInfEigenCompon{xz} = \StrainInfEigenCompon{yz} = 0}.}
%
Regarding the heat conduction, a vanishing heat flux is prescribed at the boundaries, i.e.,~\ApOne{\f{\tempFlux = \fzero \, \forall \fx \in \cVtBoundary}}
holds true.

\subsection{Heat induced by phase evolution} 
\begin{figure}[htp!]
    \centering
    \includegraphics[scale=0.65]{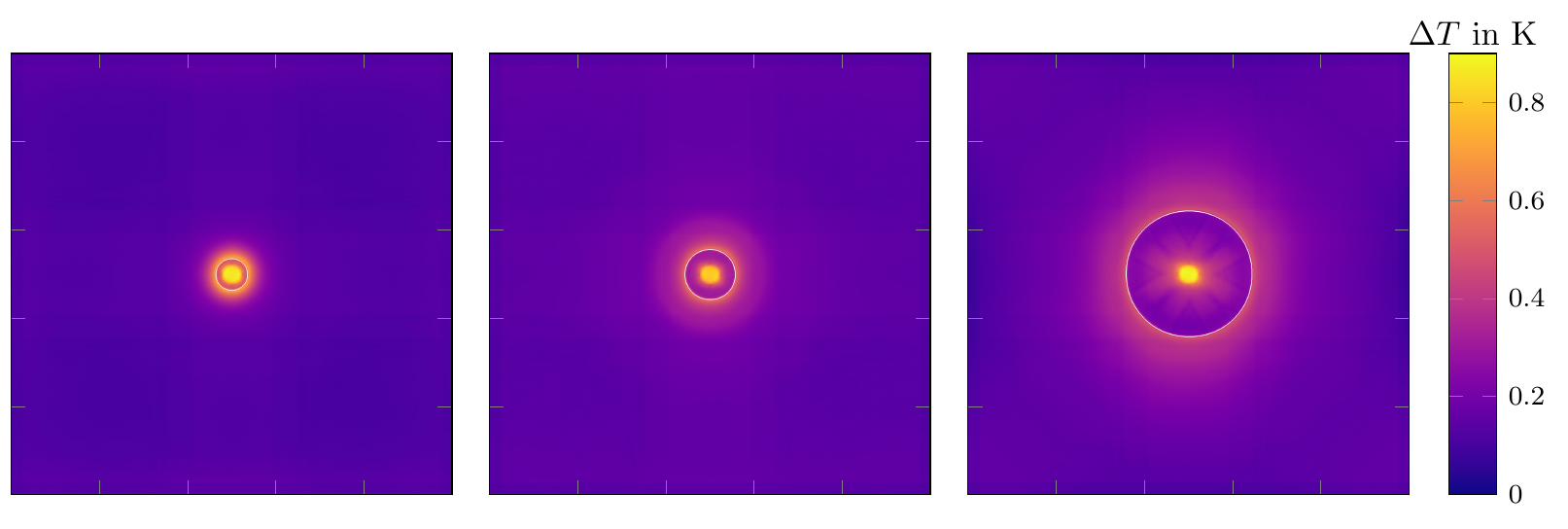}
    \caption{Temperature distribution caused by growth of inclusion and associated plastification of the matrix for three different time steps.
    The inclusion shape is indicated by the white iso-line. 
    The temperature increase is not more than~\f{1 \unit{K}} in the whole simulation domain corresponding to a representative volume element. 
    }
    \label{fig:TempDistributionByPhaseEvolution}
\end{figure}

Equation~\eqref{equation:HeatConduction_V7} indicates that both, the total and the plastic strain rate affect the evolution of the temperature. 
Thus, the plastic stress power~\f{\StressCauchy\cdot\StrainInfPlasticDot} as well as the contribution~\f{\temp \lb \pdf{\StressCauchy}{\temp} \rb \cdot \lb \StrainInfDot - \StrainInfPlasticDot \rb} due to the thermal coupling, contribute as heat source \ApOne{and} heat sink\ApOne{, respectively,} regarding the heat conduction. 
The growth of the inclusion and the thereby associated temperature distribution due to the plastification of the matrix is illustrated in~\FigRefAp{fig:TempDistributionByPhaseEvolution}.
The absolute difference of the temperature does not exceed~\f{1 \unit{K}} throughout the whole domain of the simulation\ApOne{, which illustrates a representative volume element}. 
\ApOne{The temperature increase is in the same order of magnitude as reported for the simulation of diffusionless phase transformation in~\cite{Prahs2023} and experimentally observed for TRIP steels in the context of small strains and small strain rates, \cfeg\citet[Fig. 3 \& 5]{Rusinek2009}.
}
%
\ApOne{Contrarily, the temperature increase predicted by a macroscopic continuum model in the previous section and supported by the MD simulations discussed subsequently is approximately~\f{T_{\mathrm{shock}} - T_0 \approx 103 K} according to~\TabRefAp{tab:cont_predict_temp}. 
Thus, the temperature increase due to the considered phase-transition is less than~\f{1\%} compared to the shock compression.
}
Consequently, the \ApOne{diffusionless, solid-solid} phase transformation does not provide a significant contribution to the heating with respect to the collision welding.
\ApOne{This motivates} to neglect \ApOne{this phase transition} during the impact phase of collision welding. 
\hspace{1em}


%
\section{Molecular dynamics simulation of the collision process}
\label{sec:MD}
%

From the continuum theory perspective described in Sections \ref{sec:EOS} and \ref{sec:phase_changes},  heating during high-velocity impact of a prototypical Aluminium alloy at 700 m/s results predominantly from shock compression, and plastic work. The effects of phase changes are estimated to be less than 1 K. Here, these predictions are compared to a direct simulation of the physical process using Molecular Dynamics. This yields a microstructural perspective on the collision process and allows checking the validity of the continuum theory predictions.

\subsection{Atomistic simulation at large strains}
 To investigate atomic scale processes of impact experiments, molecular dynamics simulation has shown to be a helpful tool \cite{Holian_1995}. Molecular dynamics (MD) is a numerical method at atomic scale to describe interactions between individual atoms. The technique is based on the summation of forces between single atoms derived from a local potential function. These forces can then be integrated using Newtonian laws of motion to obtain local displacements, thus giving full atomic trajectories. Due to the computation time required, only a small volume consisting of a few million atoms over some nanoseconds is usually examined. This usually implies relatively small volumes with assumed periodicity along multiple axes and generally large strain rates. This may lead to required oversimplifications for some applications, but in the present study this is well in the range of interest.

In molecular dynamics, most of the final trajectory is a result obtained without prior information, generated from just the effective force field chosen (see below). However, the physical environment in terms of statistical physics must be defined. This is done by specifying the integration ensemble and other boundary conditions.

For the present study, since the temperature distribution and pressure waves are desired result values, they cannot be controlled quantities. This leaves an isenthalpic--isochoric ensemble with an additional isobaric boundary condition perpendicular to the impact surface (but not in impact direction). This setup corresponds to the plain-stress state, with the collision axis as a free displacement boundary.

To obtain temperature and pressure profiles from these trajectories, statistical methods are applied. In MD, temperature is generally computed using a statement derived from the equipartition theorem,
\begin{equation}
    T = \frac{3}{2 \, k_B} \left< E_{kin} \right >.
\end{equation}
Here, $k_B$ is Boltzmann's constant and $\left< E_{kin} \right >$ is the ensemble average of the kinetic energy of a system in thermal equilibrium. Computing the temperature from instantaneous values is therefore not strictly valid, because the system is far away from thermal equilibrium after the massive perturbation due to the collision. However, if the ensemble average is chosen over a large enough volume and long enough time span, this conceptual issue can be neglected. In the present study, it was found that an integration region of about 20\,\AA\ with $0.1~\unit{ps}$ integration time sufficient. It is also important to compute the kinetic energy using the particular velocities after subtracting any rigid-body motion, so that only thermal velocity is considered.

\subsection{High velocity impact}
Atomistic simulation has previously been used in various high-velocity impact simulations, i.e. \cite{Bringa2006,Ecke2019,ZepedaRuiz2017}. Since the process time scales are similar to those conventionally accessible to atomistic simulation, the model assumptions are less lossy. However, care must be taken during evaluation of the resulting data.
First, the limited simulated volume often results in self-interactions. For example, a dislocation might cross over a periodic boundary and interact with its own slip plane. Similarly, elastic waves are reflected at material boundaries. In the present setup, these boundaries are the ends of the material in impact direction. This may result in the interaction of several reflections of the same wave, resulting in much higher pressure amplitudes than would appear in a bulk material where the reflecting surfaces are in the order of microseconds away. These issues are usually readily identified visually, for example by spontaneous formation of voids in solid-state cavitation.

In MD shock simulations, it is generally important to distinguish between the two different kinds of waves that can be observed. On the one hand, there are true shock waves observed as waves propagating at speeds exceeding the speed of sound in the medium. On the other hand, acoustic waves are produced by the propagation of phonons in the medium at group velocities at or below the speed of sound. This distinction is required because while the latter are nominally purely elastic and conserve enthalpy, shock waves conserve energy but do not conserve enthalpy. Macroscopically, this results in the difference between shock compression and slow decompression on the relaxation part, shown in figure \ref{fig:shock_compression_release}.
In the general case, this often leads to the splitting of a fast wave into two separate waves that propagate at different velocities, with most of the energy transferring to a wave at the speed of sound and only a small pilot wave retaining the higher velocity.

\subsection{Experimental setup}
For the atomistic simulations in this study, simulations are performed using the LAMMPS MD software \cite{Thompson2022} on the Otto von Guericke University's HPC cluster. Data post-processing and visualization is done using OVITO \cite{Stukowski2010a}.

For the simulation of a system of pure aluminium, multiple potentials were evaluated, and it was found that the MEAM potential proposed by Lee et~al. \cite{Lee2003} shows the best results: Lee's potential reproduces the $U_S$ -- $V_0$ relationship of the shock Hugoniot EOS used in Sec.~\ref{sec:EOS}. The macroscopic properties of the material described by this potential are given in \ref{tab:md_material_params}. Some of those were used in parameter fitting and are reproduced exactly, while others are implicit results obtained in validation.

\begin{table}[]
	\caption{Material parameters of aluminium as reproduced by the potential \cite{Lee2003}}
	\label{tab:md_material_params}
	\begin{tabular}{lll}
	parameter              			& value 				& units \\
	\hline
	$C_{11}$ 			            & 114.3  				& GPa     \\
	$C_{12}$ 			            & 61.9   				& GPa     \\
	$C_{44}$ 			            & 31.6  				& GPa     \\
	Calculated $B$ (VRH)            & 79.37  				& GPa     \\
	Calculated $G$ (VRH)            & 29.32  				& GPa     \\
	Calculated $E$ (VRH)            & 78.31  				& GPa     \\
	Calculated $\nu$		        & 0.336  				& -     \\
	Calculated $c_T$                & 3289                  & m/s \\
	Calculated $c_L$                & 6611                  & m/s \\
	specific volume $V_0$			& 3.69$\times 10^-4$	& m$^3$/kg \\
	Coeff. of thermal expansion     & 22.0 $\times 10^-6$   & 1 / K     \\
	heat capacity $C_P$				& 26.2					& J/(mol K) \\
                                    & 981                   & J/(kg K) \\
    melting point $T_m$             & 937                   & K \\
	heat of melting $\Delta H_m$	& 11.0					& kJ/mol \\
	\end{tabular}
\end{table}

\begin{figure}[h]
	\centering
	\includegraphics[width=8cm]{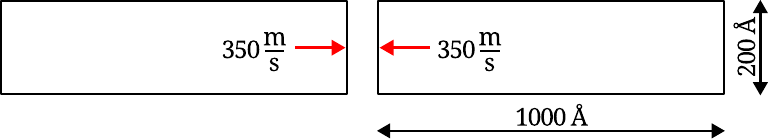}
	\caption{Setup for molecular dynamics collision simulation}
	\label{fig:md_setup}
\end{figure}

To demonstrate the capabilities, simplified collision simulations were carried out. Two pieces of material of 20\,\AA$\times$200\,\AA$\times$1000\,\AA\ of various orientations collide at a relative velocity of 700\,m/s and the resulting temperature distribution, peak temperature and welding properties are evaluated. The setup is visualised in figure \ref{fig:md_setup}. Orientations were chosen corresponding to several symmetry planes that cause no issues across the periodic boundary conditions. With respect to the global $[100]$ axis, those were $[100]$, $[110]$, $[112]$, $[134]$, with the other body in mirror symmetry.

\subsection{Results}

\begin{figure}[h]
	\centering
	\includegraphics[width=12cm]{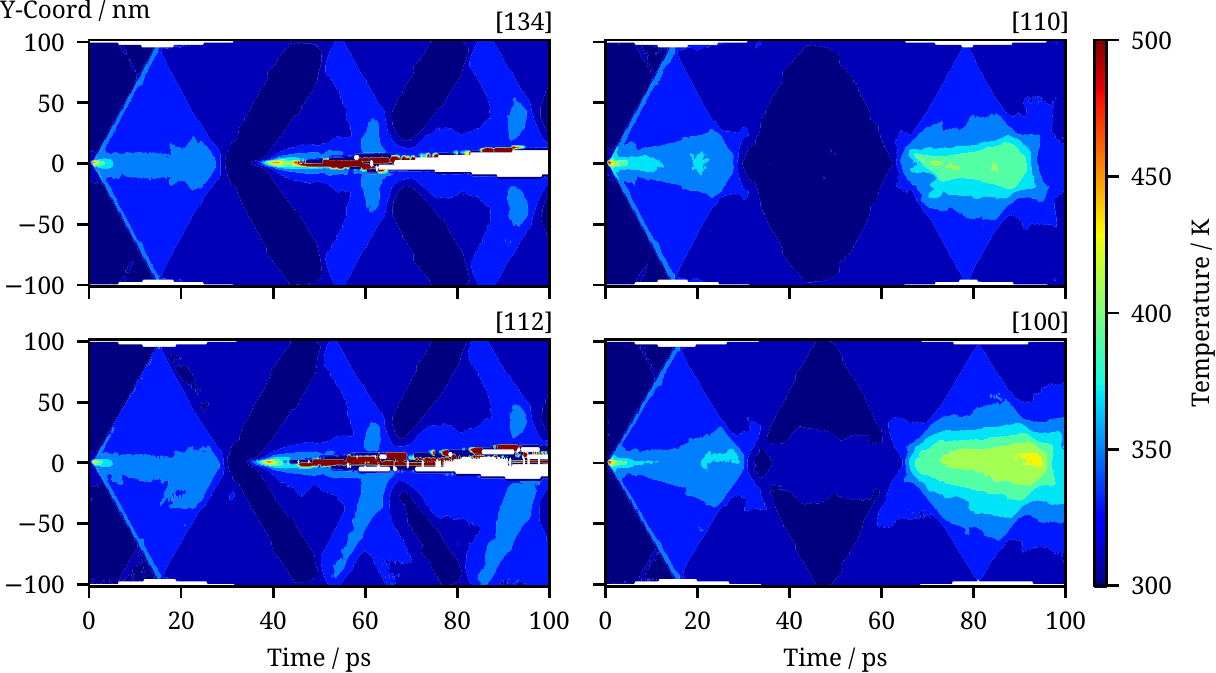}
	\caption{Temperature-time evolution of plate collision simulation. Shown are results for collision in direction $[134]$ (top left), $[122]$ (bottom left), $[110]$ (top right), $[100]$ (bottom right)}
	\label{fig:md_temp_x_t}
\end{figure}

Simulation results are summarized as a 2-D visualisation, derived by taking a temperature profile across the simulation environment's $[100]$ direction. The volume is divided into cube cells of 20 \AA $\,$ side length, for each of which the center-of-mass drift velocity is computed and subtracted from the particular velocities before obtaining the local temperature in the method described above. To increase numeric fidelity, multiple cells at the same coordinate are averaged together, resulting in one length-wise temperature profile per time step, which can be plotted as a two-dimensional representation of evolution over location and time, as shown in figure \ref{fig:md_temp_x_t}.

This form of visualisation shows several results. Diamond-like shapes are caused by the propagation of the pressure waves. All cases show wave velocities of between $6500$ and $6700~\unit{m/s}$, consistent with the longitudinal speed of sound calculated for this potential. The behaviour of the orientations [134] and [112] is completely different than what is observed for [110] and  [100]. No bonding is achieved for [134] and [112]. Instead, the collision partners bounce back after the first wave reverberation period is complete, i.e., once the compressive wave has travelled to the back surface, reflected there as a tensile wave, and come back to the collision interface. In contrast, bonding is observed for [110] and [100]. We propose that the reason for this markedly different behaviour is two-fold. First, a larger number of slip systems can be activated for [110] and [100] during collision, providing sufficient capacity for inelastic energy transfer, observed as plastic transformation. Second, the lower surface interface energy caused by better fit of the lattice orientations results in a higher resistance to rupture as the tensile waves arrive. We note that bonding observed here corresponds to pure cold welding, as the MD crystal surfaces are free from defects, impurities, or surface oxide layers as present in reality for most metals. This is different from the conditions during real collision welding, where metal surfaces with oxide layers can be successfully welded.

In the following, we focus only on the orientations where bonding occurs and evaluate the temperature evolution in a narrow band of $20~\unit{nm}$ around the weld zone with the aim of comparing against the prediction of continuum theory, i.e., $T_{\mathrm{shock}}$ and $T_{\mathrm{release}}$ as listed in Tab.~\ref{tab:cont_predict_temp}.

\begin{figure}[h]
	\centering
	\includegraphics[width=12cm]{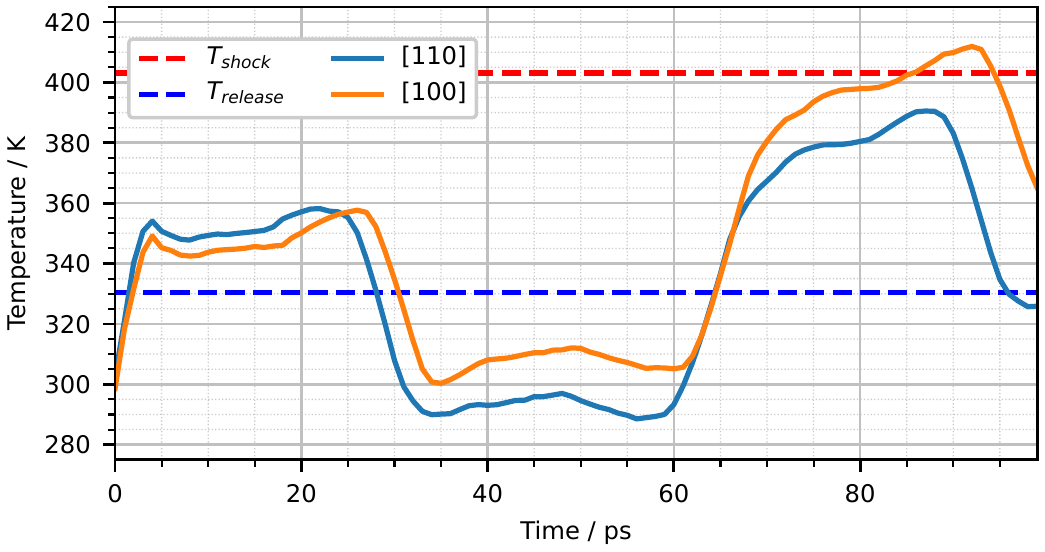}
	\caption{Temperature-time evolution of plate collision simulation for cases of successful bonding. Additionally, the predicted shock and release temperatures from table~\ref{tab:cont_predict_temp} are indicated.}
	\label{fig:md_temp_t}
\end{figure}

A plot of the temperature evolution is shown in Fig.~\ref{fig:md_temp_t}. It can be observed that the temperature changes in an oscillatory fashion, with a period corresponding to the wave reverberation time. The temperature does not immediately reach the continuum prediction $T_{\mathrm{shock}}$=403.1 K. This temperature is achieved during the second reverberation, suggesting that multiple reverberations are required to achieve sufficient exchange of potential (pressure) energy and kinetic energy to achieve a transient thermal equilibrium.

\begin{figure}[h]
	\centering
	\includegraphics[width=12cm]{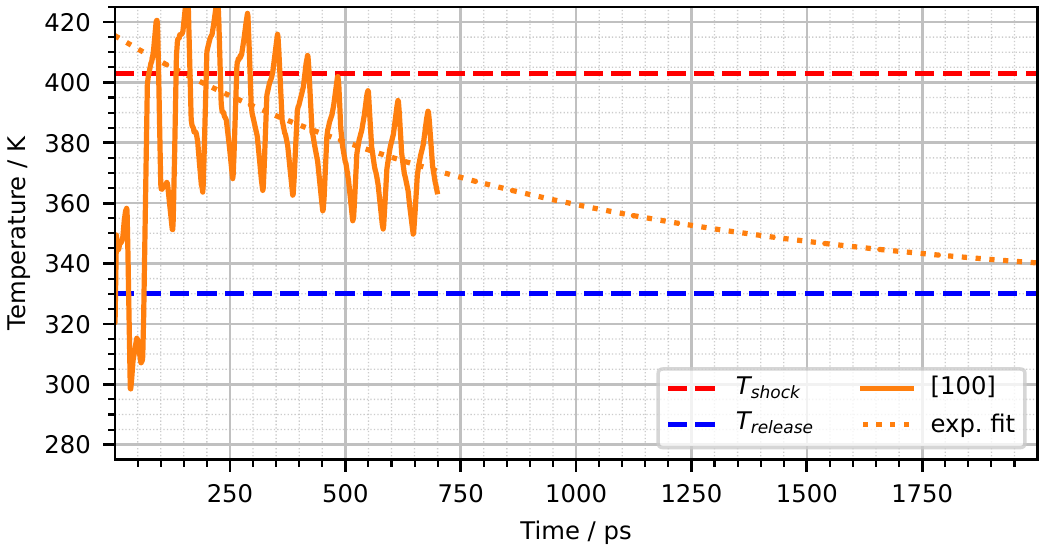}
	\caption{Temperature-time evolution of plate collision simulation for [100], simulated for a smaller system and longer time. 
 The blue solid line shows the MD data, while the dashed line shows an exponential fit towards $T_{\mathrm{release}}$. Additionally, the predicted shock and release temperatures from table~\ref{tab:cont_predict_temp} are indicated.}
	\label{fig:md_temp_long}
\end{figure}

To follow the decay of the shock compression temperature down to the continuum prediction $T_{\mathrm{release}}$ = 330.4 K, we attempted to simulate a system where bonding was successful for a longer time by using a larger proportion of the available compute resources. Fig.~\ref{fig:md_temp_long} show the temperature evolution only for the orientation [100], simulated for up to 700~ps or 10 reverberations after the initial contact. This data shows clearly that the MD results agree well for the shock compression temperature, but the simulation was still too short to fully observe decay towards $T_{\mathrm{release}}$. Nevertheless, the data is compatible with fitting an exponential decay function towards $T_{\mathrm{release}}$, suggesting that it takes approximately 50 reverberations, or 3.1~ns to approach the release temperature to within 1\%.

To summarize this section, we note that MD simulations using an MEAM potential appear to agree with the continuum theory predictions for the shock compression and subsequent release temperature. This underlines the result of Sec.~\ref{sec:phase_changes}, that the temperature contribution of phase change can be neglected. The main influence on binding appears to be the relative orientation of the crystal lattices and the resulting different energy transfer.

\clearpage

\section{Discussion and Conclusion}
This work has addressed an open issue in the context of collision welding. In this industrial process, two metal bodies are impacted onto each other at relative velocities of a few hundreds of m/s. The welding success is governed predominantly by the angle of impact, the velocity, and the mass of the metal bodies. It is currently not entirely clear which mechanisms are responsible for a successful weld. Here, we focus only the shock-induced temperature rise to quantify its contribution to melting.

To this end, the mutual impact of two aluminium plates at a relative velocity of 700 m/s at room temperature was considered. Theory on the continuum level, resting on thermodynamic concepts, predicts a temperature rise during shock compression of 103 K, and a remaining temperature rise of 30 K once the shock has been released. The physical mechanism for this temperature rise is a different compression and decompression path in thermodynamic state space, forming a dissipate hysteresis loop.
This level of theory is agnostic with respect to phase transitions. A separate treatment, based on the phase field simulation method, estimated the magnitude of temperature rise due to defect formation to be less than 1 K.

The continuum mechanics predictions were then compared with direct microstructure simulation of the process using Molecular Dynamics, resulting in a partial confirmation of the temperature rise predictions. While the transient peak temperature rise
during shock compression could be directly observed, it was impossible with the available computer resources to simulate for a sufficient time to witness the slow decay towards the release temperature. Using an exponential fit to the partial decay observed until 750 ps, it was estimated that the system decays within approximately 3 ns to the release temperature.

The agreement between continuum level theory and MD simulation results confirms that the predictions from theory are valid. Specifically no hot melt zone directly at the collision interface could be observed, a feature that continuum theory cannot resolve. Therefore, transient heating due to dynamic shock compression cannot be a dominant factor for successful collision welding. As collision welding requires the impact to occur at an angle, the welding mechanism is more likely related to the material jet which forms as a function of this angle.

\begin{acknowledgements}
\ApOne{
Andreas Prahs gratefully acknowledges the financial support of KIT excellence strategy KIT ExU-Future Fields Stage 3 ``Kadi4Mat''. 
Daniel Schneider, and Britta Nestler gratefully acknowledge the financial support of "Materials Science and Engineering (MSE)" programme No. 43.31.01, supported by the Helmholtz association. 
Andreas Prahs, Daniel Schneider, and Britta Nestler also gratefully acknowledge the support from KIT excellence strategy KIT ExU-Future Fields Stage 2 "ACDC", enabling meetings for intensive intellectual exchange on continuum thermodynamics. 
Martin Reder gratefully acknowledges partial financial support by the Bundesministerium für Bildung und Forschung (BMBF) within the KMU-innovative project BioSorb. 
Stefan Hiermaier gratefully acknowledges funding from Carl-Zeiss Foundation, grant title Materialsysteme.
}
\end{acknowledgements}



\bibliographystyle{elsart-harv}


\end{document}